\documentclass[epj,referee,epsbox]{svjour}
\usepackage{graphics}
\newcommand{\beq}{\begin{equation}} \newcommand{\eeq}{\end{equation}}
\newcommand{\beqa}{\begin{eqnarray}}    \newcommand{\eeqa}{\end{eqnarray}}
       
\newcommand{\ben}{\begin{enumerate}}    \newcommand{\een}{\end{enumerate}}
\begin{document}
\title{
Mechano-chemical coupling in growth process of actin gels and a
symmetry breaking instability } \subtitle{}
\author{
Ken Sekimoto\inst{1}\thanks{\emph{Present address:} Universit\'e
Louis Pasteur, 3 rue de l'Universit\'e, 67084, Strasbourg,
France}, Jacques Prost\inst{1}\thanks{\emph{Present address:} also
at ESPCI 10 rue Vauquelin 75231 PARIS Cedex 05, France}, Frank
J\"ulicher\inst{1}\thanks{\emph{Present address:} Max Planck
Institut f\"ur Physik komplexer Systeme N\"othnitzer Str. 38 01187
Dresden, Germany}, Hakim Boukellal\inst{1} and Anne
Bernheim-Grosswasser\inst{1}\thanks{\emph{Present address:}
Chemical Engineering Dept. Ben-Gurion University, P.O. Box 653,
84105 Beer-Sheva, Israel}
}                     
%
%
\institute{Physico-Chimie, UMR168 Institut Curie, 26, rue d'Ulm
75248 Paris Cedex 05, France}
\date{Received: date / Revised version: date}
%
\abstract{ It has been observed experimentally that
the actin gel grown from spherical beads coated with
polymerization enzymes spontaneously breaks the symmetry of its
spherical shape, and yields a ``comet'' pushing the bead forward.
We propose a mechano-chemical coupling mechanism for the
initialization of this symmetry breaking.
Key assumptions are that the dissociation of the gel takes place
mostly in the region of the external surface, and that the rates
of the dissociation depends on the tensile stress in the gel.
We analyze a simplified two-dimensional model with a circular
substrate.
Our analysis shows that the symmetric steady state is always
unstable against the inhomogeneous modulation of the thickness of
the gel layer, for any radius of the circular substrate.
We argue that this model represents the essential feature of the
three-dimensional systems for a certain range of characteristic
lengths of the modulation.
The characteristic time of the symmetry breaking process in our
model depends linearly on the radius of curvature of the substrate
surface, which
is consistent with experimental results, using spherical latex
beads as substrate.
Our analysis of the symmetry breaking phenomenon demonstrates
aspects of mechano-chemical couplings that should be working {\it
in vivo} as well as {\it in vitro}.
} 
\authorrunning{Sekimoto {\it et al.}}
\titlerunning{Mechano-chemical coupling in growing actin gels}

\maketitle
\section{Introduction}
\label{sec:introduction}
Polymerization of actin is one of the main mechanisms responsible
for
cellular motility.
Filaments of F-actin are polymerized on the cytoplasmic side of a
cellular membrane with the barbed ends oriented towards the
surface of the membrane.
The branching of the actin filaments takes place mainly
in the vicinity of the surface. 
The resulting branched F-actin filaments take a form of a soft
elastic solid, \cite{gerbal2000bis}, which we call an actin gel or
a gel, simply.
This actin gel pushes the cellular membrane  outwards.
Polymerization of actin gels is also a locomotive mechanism for
intracellular bacteria like {\it Listeria monocytogenes}, and
perhaps also for the endosomes and lysosomes (\cite{taunton2000}).
In these cases, the gel is grown in the form of a comet. This
comet pushes the bacterium forward.
For the purpose of understanding this mechanism of motility,
various experimental model systems have been developed using both
biochemical and biophysical approaches.

Biochemical approaches have isolated 
the basic cytoplasmic ingredients needed for the motility
of {\it Listeria monocytogenes} \cite{loisel1999}:
(1) {actin and ATP} for the formation of F-actin filaments, (2)
Arp2/3 as the cross-linker and/or the nucleator of the F-actin
growth (i.e., the precise role is still under debate), (3) ADF as
the depolymerization factor at the pointed end of F-actin,
(4) the capping protein,
(5) a bacterial protein called ActA expressed on the surface of
{\it Listeria} which is necessary for inducing polymerization from
the surface.
These ingredients constitute a model 
cytoplasm for the motility.
\\

Biophysical approaches have taken the {\it Listeria} as a model
system of cellular motility.
Furthermore, a bio-mimetic {\it in vitro} system of the bacterial
motility has been introduced. This system consists of
a spherical latex bead, 
coated by the enzymatic protein complexes, ActA (
\cite{cameron1999}, \cite{noireaux2000}),
or a fragment of its homologue from human cells, WASP
[{Wiskott-}{Aldrich-}{Syndrome} Protein]
\cite{yarar1999,anne2002}.
The cytoplasm has also been replaced by the reconstituted
cytoplasm \cite{anne2002}.
Despite the spherical form of the bead, the gel has grown in a
shape of a comet, like the bacteria {\it Listeria} (see for
example Fig.2 of \cite{yarar1999}).
{\bf Fig.\ref{fig:1}} 
shows the initial stage of the creation of
the comet, observed using fluorescent probe attached to actin monomers.\\
\begin{figure*}  
\vspace*{5cm}       
\caption{Experimentally observed time sequence of the actin gel
grown around a spherical latex bead. The diameter of the bead is
10 $\mu$m.
The bead is coated with 
a fragment of WASP, and it is placed in a reconstituted
solvent as 
described in the text. The actin monomers in the gel is visible by
a fluorescent marker.
The observation started at $t=0$ after several tens of seconds
when gel has started to grow.}
\label{fig:1}       
\end{figure*}
This process bares the signature of a spontaneous symmetry
breaking, which is the subject of the present paper.
The phenomenon of the symmetry breaking is relevant to some
biological systems of sub-cellular level.
For instance the endosomes, which consist of spherical soft
substrate (liquid vesicle), grow a comet \cite{taunton2000}.
Also, the motility of a mutant {\it Listeria}, which moves
preferentially in lateral directions
\cite{rafelski2002} grows the actin gel by %
breaking its cylindrical symmetry.
Our principal aim is to assess, through the study of the symmetry
breaking, the relevance of the elastic aspects to the biological
motilities based on the polymerization of protein filaments, and
to provide for several basic ingredients related to the
mechano-chemical coupling.
The three ingredients essential for explaining the symmetry
breaking are (details will be given in sections \S 2-\S 4):

{\it (i) The creation of a tensile stress due to the curvature of
the
substrate surface} (\S\ref{sec:symmetric}):\\
 As the gel is continuously created at the bead surface
(at the radius, $r=r_0$), the part that has been already formed is
continuously pushed outwards ($r>r_0$). Since the perimeter ($2\pi
r$) increases as $r$, and since the surface has a closed topology,
the gel is stretched by the ratio, $r/r_0 (>1)$.

{\it (ii) The concentration of the tensile stress by a geometrical
effect} (\S\ref{sec:concentration}): \\ 
The gel layer around a bead is in mechanical equilibrium, so that
the integrated tension across the layer thickness of the gel must
be constant along the surface of the bead. In particular, if the
thickness is locally thinner, such a region must bear a stronger
tension in order to support the same integrated tension.

{\it (iii) The acceleration of dissociation of the gel
under tensile stress} (\S\ref{sec:mechano_chem}):\\
We suppose that, under tensile stress, the gel dissociation is
accelerated through the mechano-chemical coupling. This
dissociation may be either through the unbinding of the branching
points along actin filaments, or through the depolymerization of
actin filaments.

These three ingredients {\it (i)}-{\it (iii)} constitute a
positive feedback loop leading to an instability of the symmetric
shape of the growing gel. This will be described in
\S\ref{subsec:evolution}. In short, the region of gel with smaller
thickness becomes preferentially dissociated due to the higher
tensile stress, implying further thinning of that region.

%
All symmetry breaking models \cite{oudenaarden1999,mogilner2003}
take the mechano-chemical coupling into account. Previous models
have focused their attention on the compressive force acting on
the actin filaments at the polymerization sites, that is on the
substrate surface.
Our analysis takes into account the global stress distribution.
Of particular importance is the tensile stress generated at the
outer surface of the gel.
Indeed, on general grounds, the depolymerization rate must be an
increasing function of the tensile stress.
We show in the following that it leads inevitably to symmetry
breaking.
In the discussion section \S\ref{sec:discussion}, after a brief
summary, we compare, in more details our analysis with the
existing ones, and suggest experiments designed for distinguishing
between the different possibilities.
%
%
%

\section{Distribution of stress within the gel with
symmetric shapes}%
\label{sec:symmetric}

Suppose that a gel has been polymerized steadily from a substrate
surface of either spherical or cylindrical shape with radius
$r_0$, until the gel forms a layer of a thickness $h$, enclosing
the substrate surface and keeping its symmetry
(see, {\bf Fig.\ref{fig:2}}).
\begin{figure} 
\vspace{5cm} %
\caption{The cross section of F-actin gel around a bead. The
relevant stress components in the gel are schematically shown. The
gel occupies the space between the radii $r=r_0$ and $r=r_0+h$.
The compressive component of the stress at the substrate surface
($r=r_0$), $\left.\sigma_{rr}\right|_{r_0}$,  and the tensile
component at the outer surface ($r=r_0+h$),
$\left.\sigma_{\perp\perp}\right|_{r_0+h}$, are indicated by the
pairs of oppositely oriented open arrows.
In this symmetric state, the tensile component at the substrate
surface, $\left.\sigma_{\perp\perp}\right|_{r_0}$,
 as well as the normal compressive
component at the outer surface,
$\left.\sigma_{rr}\right|_{r_0+h}$, vanish.} %
\label{fig:2} %
\end{figure} %
As already noted, the part of the gel that has been formed has
been continuously pushed outwards.
An element of the gel at the radius $r$ is then stretched by
$r/r_0$ times relative to the native state of polymerization.

%
To know the tensile stress in the lateral direction,
$\sigma_{\perp\perp}$, let us use the ``stacked rubber band
model'' \cite{noireaux2000,gerbal2000,landau1967}:
A freshly cross-linked gel at the latex surface ($r=r_0$) is
unstretched and has no lateral stresses,
$\left.\sigma_{\perp\perp}\right|_{r=r_0}=0.$
As the layer is pushed outward, its circumference increases which
introduces a lateral stress (\cite{gerbal2000}), \beq \left.
\sigma_{\perp\perp} \right|_{{r}} = B\frac{r-r_0}{r_0},
\label{eq:rubber} \eeq with $B$ being the Young modulus.
In particular, when  the thickness of the gel layer is $h$, the
tangential stress at the outer surface of the gel is \beq
\sigma_{\perp\perp}\vert_{r_{0}+h} = B \frac{h}{r_{0}}.
\label{eq:rub_exterior} \eeq
We remark that the present approximation ignores the radial
deformation due to the lateral stretching, in other words, it
assumes a vanishing Poisson ratio. Although this has no
justification for actin gels, the main results of the present
paper do not depend on this property. More refined calculation
confirm the validity of this statement \cite{sekimoto2001}.

The shear component of the stress $\sigma_{r\perp}$ vanishes
everywhere, for symmetry reasons:
\[\sigma_{r\perp}=0.\]
Radial force balance requires that
 the radial stress, $\sigma_{rr}$,
at the radius $r$ obeys the following equations: \beq
\frac{1}{r^2}\frac{\partial}{\partial r}\left( r^2 \sigma_{rr}
\right) -\frac{2}{r}\sigma_{\perp\perp} =0 \label{eq:bl3} \eeq for
a spherical surface (\cite{noireaux2000}, see
 Appendix A), or
\beq \frac{\partial}{\partial r}\left(r \sigma_{rr} \right)
-\sigma_{\perp\perp} =0 \label{eq:bl2} \eeq for a cylindrical
surface \cite{gerbal2000}.
%
%
Since no external force is applied on the outer surface of the gel
 layer,
the normal stress must vanish: \beq \left. \sigma_{rr}
\right|_{{r_0}+h}  = 0. \eeq
Under this condition, the normal stress at the substrate surface,
$\left.\sigma_{rr}\right|_{r_0}$, can be calculated in terms of
the lateral stress, $\sigma_{\perp\perp}$. In the case of
cylindrical substrate, we integrate Eq.(\ref{eq:bl2}) from $r=r_0$
to $r=r_0+h$, and have \beq \left.\sigma_{rr}\right|_{r_0}=
-\frac{T}{r_0}, \label{eq:laplace} \eeq where $T$ is the {\it
integrated tension} across the symmetric gel slab, defined by \beq
T=  \int_{{r_0}}^{{r_0}+h} \sigma_{\perp\perp}dr. \label{eq:T}
\eeq

Using Eq.(\ref{eq:rubber}) we find $T=B h^2/(2r_{0})$, and thus
$\sigma_{rr}\vert_{r_{0}}=-B h^2/(2r_{0}^2)$.
For the spherical substrate, the relation is not as simple as the
cylindrical case.
Still, $\sigma_{rr}$ is given as an integration of
$\sigma_{\perp\perp}$.

\section{Concentration of the tensile stress under a
modulated surface profile} \label{sec:concentration}
In this section, we consider how small perturbations to the
surface profile of the gel layer lead to the redistribution of the
stress components within the gel layer.
{   
We introduce the function representing the thickness of the gel
layer, $h(\hat{\omega})$, with the variable $\hat{\omega}$
representing the orientation from the origin.
A spherically symmetric gel layer corresponds to the constant
function, $h(\hat{\omega})=h^*$, with a constant thickness $h^*$.

The analysis of the thickness perturbation
 is done in the following two steps:
In the first step, we suppose that this function,
$h(\hat{\omega})$, is slightly perturbed from a constant function,
but we still do not allow for the displacement of the gel.
%
%
In the second step, we let the gel layer  relax until it
reestablishes the mechanical balance.
We calculate how the stress in the gel is distributed in this new
balanced state.
To avoid any confusion, we stress that, the perturbations
($h(\hat{\omega})-h^* \neq 0$) at the end of the first step does
{\it not} imply the swelling or deswelling of the gel layer.
The perturbation rather implies that there is more or less
material of gel along the direction $\hat{\omega}$ than the
average.
(It could be due to the enhanced/depressed polymerization, or, to
the depressed/enhanced dissociation of the gel along this
direction.)
As we discuss a situation such that there is
 no external force on the outer surface,
we require stress-free conditions on the outer surface of the gel
layer.
}   
\beq \left.\sigma_{rr}\right|_{r_0+h}
  =\left.\sigma_{r\perp}\right|_{r_0+h}=0.
\label{eq:stressfree} \eeq

What we demonstrate is that, under the above conditions, the
tensile stress $\sigma_{\perp\perp}$ under the reestablished
mechanical balance
{   
is locally enhanced in the thinned region of the gel layer, that
is, in the zone of the orientation $\hat{\omega}$ that satisfies
$h(\hat{\omega})<h^*$.
}   
On the one hand, the physical origin of the stress concentration
is quite simple and universal.
In fact the authors have noticed, after completion of the present
work, that essentially the same mechanism of stress concentration
has been discussed long before in the context of crystal growth
under stress (see, for example, a concise review on the related
history in the literature \cite{kassner2001}).
In Appendix B we describe the basic mechanism of this phenomenon
by using an illustrating example in a very simple geometry.
On the other hand, the direct analysis of the present case with
the distributed thickness $h(\hat{\omega})$ is difficult, because
of the three spatial dimensionalities and the tensorial character
of the stress associated to this space.
We can avoid, however, this difficulty by
the following lines of reasoning.\\
\indent {\it 1.} We limit our concern to the modulations
$h(\hat{\omega})-h^*$ whose characteristic wavelengths are
comparable to the average thickness, $h^*$.
Experimentally, $h^*\!/r_0$ is at most about 0.2
\cite{gerbal1999}. The radius of curvature of the outer surface
($\simeq r_0+h^*$) is, therefore, not appreciable in view of such
short wavelength of modulation.
We may then ignore the effect of a specific curved geometry of the
substrate surface {\it except for} the fact that the curvature
gives rise to the lateral tension
$\sigma_{\perp\perp}$ in the gel layer.\\
\indent {\it 2.} We notice the following fact: As far as the
stress distribution inside the gel layer is concerned, the
influence of the surface profile perturbation is practically
limited to a region near the outer surface (see, {\bf
Fig.\ref{fig:3}}).
\begin{figure} 
\vspace{5cm} %
\caption{Schematic representation of the outer gel surface before
(a) and after (b) perturbation. A perturbation of the surface
profile with characteristic length $\lambda$ affects the stress
profile only within a ``skin layer'' of thickness
$\lambda$.} %
\label{fig:3} %
\end{figure} %
More precisely, if the perturbation is characterized by a
wavelength, $\lambda$, then the thickness of the disturbed region
 is also characterized by $\lambda$.
(The boundary condition far from this layer is therefore
irrelevant to this disturbance.)
For detailed arguments, see Appendix C.
%
%

\indent {\it 3.} { With our wavelength choice in {\it 1.}, and
with the fact just mentioned above {\it 2.},
we can justify the study of {\it (a)} a two dimensional circular
geometry rather than the real spherical one, with {\it (b)} a
``slip'' boundary condition on the substrate surface, to see how
the stress in the actual three-dimensional case is distributed
after the reestablishment of the mechanical balance.
Moreover, {\it (c)} the neglect of the shear stress components
within the gel layer is justifiable for the experimentally
realized situation where the mean thickness of the gel layer $h^*$
is much smaller than the radius $r_0$.
We will formulate these assumptions in more details below:\\
}   
{\it (a) We consider the gel layer grown around a two dimensional
circle of radius $r_0$.} \\
In two dimension, we represent the thickness of the gel layer by
$h(\theta_0)$ as a function of the angle $\theta_0$ with $0\le
\theta_0 <2\pi$, instead of $h(\hat{\omega})$ above. (See, {\bf
Fig.\ref{fig:4}}: $h(\theta_0)$ is defined {\it before} the
reestablishment of the mechanical balance.)
\begin{figure} 
\vspace{5cm} %
\caption{Definition of the angler variable $\theta_0$ and the
hight function $h(\theta_0)$ of a gel with modulated thickness due
to depolymerization before an elastic deformation reestablishes
mechanical equilibrium.
} %
\label{fig:4} %
\end{figure} %
The lateral components of the stress, which we have denoted
symbolically by $\perp$, corresponds now to the azimuthal
direction. We then use the suffix $\theta$ in place of $\perp$
hereafter. For example, we write $\sigma_{r\theta}$ for
$\sigma_{r\perp}$, and $\sigma_{\theta\theta}$ instead of
$\sigma_{\perp\perp}$.
For small perturbations of the thickness, $|h(\theta_0)-h^*|/h^*
\ll 1$, we may use the linear analysis.
Then it suffices to consider the form \beq h(\theta_0)= h^*\left[1
+ \epsilon_q \cos (q \,\theta_0 )\right], \label{eq:perturbation}
\eeq
where the integer $q$ indicates the number of nodes of the spatial
undulations, and $\epsilon_q$  is supposed to be small
($|\epsilon_q|\ll 1$).
The characteristic wavelength for the the $q$-th mode is about $
2\pi r_0/q$, and the restriction (1) is represented as
$q \simeq \frac{2\pi r_0}{h^*}$.\\
%
(Remark: Besides our purpose of analysis, the two-dimensional
geometry applies rather directly to a  {\it Listeria} mutant
\cite{rafelski2002} mentioned in \S\ref{sec:introduction}:
This mutant moves preferentially in lateral
directions, breaking its cylindrical symmetry.) \\
\noindent {\it (b) On the substrate surface ($r=r_0$),
the shear stress is negligible.} \\
The slip boundary condition for the shear stress is written as
\beq \left.\sigma_{r\theta}\right|_{r_0}=0 \label{eq:slip}.
\label{eq:cisaisub} \eeq
({Remark: Note that we do not claim this boundary condition to be
always realistic. We rather use this condition since it is
justifiable for the calculation of the stress distribution under
the modes of perturbations with $q\simeq \frac{2\pi r_0}{h^*}$:
See the argument {\it 2} above and the appendix C for the details.
})\\
{\it (c) The shear stress $\sigma_{r\theta}$ within
the gel layer is negligible.}\\
As mentioned above, the experimental value of $h^*\!/r_0$ is $\ll
1$.
In such situation we may, in the lowest order approximation,
estimate the magnitude of the shear stress, with a parabolic
profile of the shear stress $\sigma_{r\theta}$:
$\sigma_{r\theta}=\tilde{\mu}\epsilon_q (r_0+h-r)(r-r_0)/r_{0}^2 $
for $r_0\le r\le r_0+h$, which satisfies the boundary conditions,
Eqs.(\ref{eq:stressfree}) and (\ref{eq:cisaisub}). Here,
$\tilde{\mu}$ is a constant proportional to the shear modulus
$\mu$ of the gel.
The magnitude of $\sigma_{r\theta}$ is, therefore, at most of the
order of $\tilde{\mu}\epsilon_q {(h^*\!/r_0)}^2 $.
We compare this with the change of $\sigma_{\theta\theta}$ due to
the perturbations of the thickness, which is of order $\epsilon_q
B h^*\!/r_0$ (see Eq.(\ref{eq:rubber})).
Then $\sigma_{r\theta}$ is smaller than this by a factor of
$h^*\!/r_0$, and is therefore negligible.

In the Appendix D we show how the stress distribution within the
gel is calculated for the model described by {\it (a)}-{\it (c)}.
%
Below we show only the results for the tensile stresses
$\left.\sigma_{\theta\theta}\right|_{r_0+h}$ at the external gel
surface and the normal compression,
$-\left.\sigma_{rr}\right|_{r_0}$ at the substrate surface:
\beq \left.\sigma_{\theta\theta}\right|_{r_0+h}
-\left(\left.\sigma_{\theta\theta}\right|_{r_0+h}\right)_{\epsilon_q=0}
  = -  \frac{ B\chi^2 }{2+\chi }
\epsilon_q \cos ( q\, \theta_0 )+{\cal O}({\epsilon_q}^2)
\label{eq:lateral-lin} \eeq
\beq
 \left. -\sigma_{rr} \right|_{{r_0}}
-\left( \left. -\sigma_{rr} \right|_{{r_0}} \right)_{\epsilon_q=0}
= {\cal O}({\epsilon_q}^2), \label{eq:sigma_rr} \eeq
with $\chi \equiv h^*\!/r_0$. The bracketed terms with the
subscript $\epsilon_q=0$ are those terms without perturbation:\\
$\left(\left.\sigma_{\theta\theta}\right|_{r_0+h}\right)_{\epsilon_q=0}
=B\chi$ and $\left( \left. -\sigma_{rr} \right|_{{r_0}}
\right)_{\epsilon_q=0} =\frac{B}{2} \chi ^2$ (see
Eqs.(\ref{eq:rub_exterior}) and (\ref{eq:laplace})). In
Eq.(\ref{eq:sigma_rr}), ${\cal O}({\epsilon_q}^2)$ indicates the
terms of at least second order of ${\epsilon_q}$.
Since $\epsilon_q \cos (q \,\theta_0 )=(h(\theta_0)- h^*)/h^* $,
the minus sign on the right hand side of Eq.(\ref{eq:lateral-lin})
implies that the lateral tension is augmented, $ \left.
\sigma_{\theta\theta} \right|_{{r_0}} > \left( \left.
\sigma_{\theta\theta} \right|_{{r_0}} \right)_{\epsilon_q=0} $, in
the thinned portion of the layer, $h(\theta_0)<h^*$.
%

\section{Mechano-chemical coupling:
Growth and dissociation of gel under stress}
\label{sec:mechano_chem}

In this section, we consider the time evolution of the thickness
of the gel layer.
We denote the profile of the thickness at the time $t$ as
$h(\theta_0,t)$.
We are interested in the chemical processes which take place on
time scales much larger than the establishment of the mechanical
balance within the gel.
We suppose that the relevant microscopic chemical processes are
the polymerization and branching of the actin filaments to form
the gel, and the unbinding of the branching points and/or through
the depolymerization of actin filaments to dissociate the gel.
We adopt a simplified version of the model proposed previously
\cite{gerbal1999,noireaux2000}:
\beq \frac{\partial h(\theta_0,t)}{\partial t}= a \left[ \bar{k}_p
e^{ \left. \sigma_{rr} \right|_{{r_0}}c_p } \,-\, {k}_d e^{
\left.\sigma_{\theta\theta}\right|_{r_0+h(\theta_0,t)} c_d}
\right], \label{eq:reac0} \eeq where $a$, $\bar{k}_p$, $k_d$,
$c_p$ and $c_d$ are positive constants.
The prefactor $a$ outside the square bracket on the right hand
side (r.h.s.) is a length of about the size of an actin monomer.
This represents the rate of conversion between the chemical
processes and the change of the thickness, $h$.
The other parameters are described below.

In the square bracket on the right hand side (r.h.s.) of
Eq.\ref{eq:reac0}, the first term represents the polymerization at
the substrate surface ($r={r_0}$).
Here we have introduced the assumption: {\it (i) On the substrate
surface, the polymerization is the dominant process.}
The pre-exponential factor $\bar{k}_p$ represents the
kinetic constants in the 
absence of compressive stress ($\left.\sigma_{rr}
\right|_{{r_0}}=0$).
$\bar{k}_p$ depends on the concentration of actin monomers in the
solvent. In our analysis we assume this to be constant.
The exponential factor represents the fact that the polymerization
is decelerated by the compression, $ \left.\sigma_{rr}
\right|_{{r_0}}$  $(<0)$.
The parameter $c_p$ has been introduced so that
 $ - c_p \left.\sigma_{rr} \right|_{{r_0}}$
accounts for the increase in the polymerization potential barrier
due to the cost in elastic energy (divided by $k_{\rm B} T$) to
push out the gel layer outward against the compressive stress.
We have neglected the dissociation of the gel at the substrate
surface.
Such process could be easily incorporated in the model
\cite{prost2001,sekimoto2001}, but has little effect in our
context.
In the experiment of the polymerization of microtubules, it has
been shown that the negligence of the depolymerization on the
growing end (the plus end) is a good approximation
\cite{dogterom1997}.
%

The second term in the square bracket on the r.h.s. of
Eq.\ref{eq:reac0} represents
the gel dissociation. 
We have introduced the assumption: {\it (ii) The dissociation
process is almost localized on the outer surface of the gel at
$r=r_0+h(\theta_0,t)$.}
%
The pre-exponential factor, $-a k_d$, therefore represents the
rate of thickness decrease which occurs due to the dissociation of
the gel under the  stress-free condition,
$\left.\sigma_{\theta\theta}\right|_{r_0+h(\theta_0,t)}=0$.
(Remark: Here we can identify
$\left.\sigma_{\theta\theta}\right|_{r_0+h(\theta_0,t)}$ as the
tensile stress along the tangent of the outer surface, since the
correction is of second order of the deviation angle,
$|\frac{\partial h}{\partial \theta_0}|/(r_0+h)$, between the
tangential direction and the azimuthal direction.)
The exponential factor of this term represents the fact that the
dissociation is accelerated by the lateral tensile stress
$\left.\sigma_{\theta\theta}\right|_{r_0+h(\theta_0,t)}$  $(>0)$.
The parameter $c_d$ has been introduced so that
 $  c_d \left.\sigma_{\theta\theta}\right|_{r_0+h(\theta_0,t)}$
accounts for decrease in the depolymerization potential barrier
due to the release of the elastic energy (divided by $k_{\rm B}
T$) when the gel is dissociated under the tensile stress.
We have neglected the dissociation of the gel occurring
inside the gel. %
There are good reasons to belive that the gel dissociation is
strongly accelerated under tensile stress \cite{prost2001}, as
compared with spontaneous dissociation under the stress-free
condition.
%
{   
In fact, the experiments using the full cell extract as the
solvent have shown that the mean thickness of the gel layer around
the latex bead is much smaller than the average length of the
comet produced by {\it Listeria} of similar size.
}   
It implies that the Boltzmann factor of the form, $e^{c_d
\sigma_{\theta\theta}}$, is crucial to determine the dissociation
rate.
As $\sigma_{\theta\theta}$ is largest on the outer surface of the
gel, we suppose that the gel dissociation occurs mostly in the
vicinity of the outer surface.

The kinetic equation Eq.(\ref{eq:reac0}) also assumes the
following: {\it (iii) The diffusion of actin monomer is fast
enough.}
This limits our analysis to a  bead radius range smaller than a
cross-over size, $r_c$, separating a stress governed regime from a
diffusion controlled regime.
Indeed, on the substrate surface, the actin gel is formed from the
adjunction of actin monomer molecules.
And for these molecules to reach the substrate surface, they have
to diffuse through the network of the actin gel.
Previous experimental and theoretical analysis \cite{noireaux2000}
indicates that, as far as the diameter of the latex bead is less
than about 5 $\mu$m, and under physiological concentrations of the
actin monomers and of the cross-linker molecules, 
 diffusion does not limit the thickness evolution,
 $h(\theta_0,t)$.
%

The evolution equation Eq.(\ref{eq:reac0}) has a solution
corresponding to the symmetric stationary state,
$h(\theta_0,t)=h^*$ \cite{noireaux2000}.
 If we restrict our
analysis to
 the circularly symmetric profiles,
$h(\theta_0,t)=h(t)$, this solution is {\it stable}.
In fact, substituting the form $h(\theta_0,t)=h^*$ into
Eq.(\ref{eq:reac0}), we obtain the equation for $\chi \equiv
h^*/r_0$ as
\beq \frac{c_p}{c_d}{{\chi }}^2 + {{\chi }} -\frac{2}{c_d B}
\log\left(\frac{\bar{k}_p}{k_d}\right)=0. \label{eq:chi} \eeq
This equation has a positive, therefore physically meaningful,
 solution for $\bar{k}_p/k_d>1$,
Furthermore, if we substitute the form
\beq h(\theta_0,t)= h^*\left[1 + \epsilon_0(t) \right],
\label{eq:perturbation0} \eeq
the Eq.(\ref{eq:reac0}) reduces, up to linear order in
$\epsilon_0(t)$, to the following equation:
\beq \frac{d\epsilon_0(t)}{dt} = -\frac{\epsilon_0(t)}{\tau_0},
\label{eq:epsilon0} \eeq with $\tau_0= {k_d}^{-1} {\Omega_0}^{-1}
r_0/a$, \beq \Omega_0= c_d B    e^{c_d B {\chi }}
\left(1+\frac{c_p}{c_d}{\chi }\right). \label{eq:omega0} \eeq
Equation (\ref{eq:epsilon0}) shows, as already mentioned, that the
steady state solution $h(\theta_0,t)= h^*$ is stable with respect
to perturbations keeping the overall symmetry.
This result is understandable since a radius with
 $h>h^*$ ($<h^*$) would lead to an increase (decrease) of
both $( -\left.\sigma_{rr} \right|_{{r_0}}) $ and
$\left.\sigma_{\perp\perp}\right|_{r_0+h}$, and these in turn make
the r.h.s of Eq.(\ref{eq:reac0}) negative (positive), leading to a
decrease (increase) of $h$ toward the stationary value $h^*$.
From Eqs.(\ref{eq:chi}) and (\ref{eq:omega0}), $\chi $ and
$\tau_0$ are functions of three parameters, $c_d B
,\frac{c_p}{c_d},$ and $\frac{\bar{k}_p}{k_d}.$
%
Note that $\tau_0$ is a few orders of magnitude larger than the
microscopic time ${k_d}^{-1}$, with $r_0/a$
 being of order $10^3$ and $\Omega_0$ of order 10.

\section{Result}
\subsection{Symmetry breaking instability}
\label{subsec:evolution}

We now consider symmetry breaking perturbations and we assume the
following form for the gel layer profile:
\beq h(\theta_0,t)= h^*\left[1 + \epsilon_q(t) \cos (q \,\theta_0
)\right], \label{eq:perturbationt} \eeq
with $q\neq 0$.
%
Substituting the expressions of the stress components,
Eqs.(\ref{eq:lateral-lin}) and (\ref{eq:sigma_rr}) into
Eq.(\ref{eq:reac0}), where $\epsilon_q$ is replaced by
$\epsilon_q(t)$, we have the following equation up to the linear
order of $\epsilon_q(t)$,
\beq \frac{d\epsilon_q(t)}{dt}= \frac{\epsilon_q(t)}{\tau_q},
\label{eq:linear} \eeq with $\tau_q =  \tau_0
\frac{\Omega_0}{\Omega_q},$ \beq \Omega_q = c_d B \, e^{c_d B
{\chi }} \frac{{\chi }}{2+{\chi }}, \label{eq:omegaq} \eeq where
$\chi \equiv h^*/r_0$ as before. (Remember that $\chi$ can be
expressed in terms of the parameters
$c_d B ,\frac{c_p}{c_d},$ and $\frac{\bar{k}_p}{k_d}.$)\\
%
Note that since ${\Omega_0}{\Omega_q}\simeq 2/\chi \simeq 10$,
 $\tau_0 \ll \tau_q$.
Eqs.(\ref{eq:linear}) and (\ref{eq:omegaq}) imply the following
characteristics of the symmetric stationary state
$h(\theta_0,t)=h^*$.\\
{\it (i) The symmetric stationary state is unstable against
perturbations which break the symmetry,} since all $\tau_q$ are
positive.
In fact, the applicability of our model
is guaranteed only in the range of $q$ satisfying $q \simeq
\frac{2\pi r_0}{h^*}$ (see, \S\ref{sec:concentration}).
Nevertheless, the presence of {\it an} unstable mode is sufficient
for the proof of instability.
Note also that since $\tau_0 \ll \tau_q$, our analysis predicts
that a quasi symmetric steady state should be reached
significantly earlier than the onset of symmetry breaking. This is
indeed what is observed.
\\
{\it (ii) The characteristic time of the instability is
proportional to the radius of the substrate if the other
parameters are fixed.}: It is reasonable to suppose that $\tau_q$
represents the characteristic time of the growth of the
perturbation.
Then, from (\ref{eq:linear}), $\tau_q$ is written in a scaling
form: \beq \frac{\tau_q}{{k_d}^{-1}} ={\Omega_q}^{-1}\,
\frac{r_0}{a}, \label{eq:tau} \eeq where ${k_d}^{-1}$ and $a$ play
the role of intrinsic timescale and lengthscale, respectively.
The dimensionless constant of proportionality, ${\Omega_q}^{-1}$,
depends on the properties of the gel and of the solvent through
the parameters, $c_d B ,\frac{c_p}{c_d},$ and
$\frac{\bar{k}_p}{k_d}$.
Note that, in fact, the quantity $\tau_q$ thus defined shows no
dependence on $q(\neq 0)$, as $\Omega_q$ does not.
This apparently anomalous behaviour should not be taken seriously,
because the range of wavenumber validity of of our analysis is
limited to $q \simeq \frac{2\pi r_0}{h^*}$.)

Quantitatively, we can evaluate the characteristic time $\tau_q$
using the experimentally known data in the literature:
The stationary velocity $v_{\rm gel}$ at which the gel material
moves outward is identified from Eq.(\ref{eq:reac0}) as\\ $v_{\rm
gel}/r_0=$ $ {k}_d ({a}/{r_0}) e^{c_d
\left.\sigma_{\theta\theta}\right|_{r_0+h(\theta_0,t)}} ={k}_d
({a}/{r_0}) e^{c_d B\chi}$.
Comparing this with the  expression of $\tau_q$ obtained from
Eqs.(\ref{eq:omegaq}) and (\ref{eq:tau}), ${\tau}^{-1}=$ $k_d
({a}/{r_0})$ $e^{c_d B {\chi }}$ $c_d B {{\chi }}/({2+{\chi }})$,
we see that
\beq \tau=\frac{r_0}{v_{\rm gel}}\frac{2+\chi}{c_d B\chi}.
\label{eq:taur0} \eeq
%
As described in \cite{noireaux2000}, $c_d B \chi$ is the decrease
in energy barrier (in units of $k_{\rm B}T$) in the dissociation
of an actin filament under tensile stress, compared to the
unstressed case.
For the effects described in this manuscript to be observable,
this decrease must be of order one. Noting that $\chi \ll 2$, our
analysis requires the combination $\tau_q v_{\rm gel}/(2r_0)$ to
be of order one.
The experiment gives $\tau_{\rm sym}/r_0 \simeq 5$ min/$\mu$m and
$v_{\rm gel}\simeq 1 \mu$m/min (note that it is the polymerization
rate under stress), which leads to
 $\tau_q v_{\rm gel}/(2r_0)\simeq 2.5$. This is in the expected
range.
%

\subsection{Role of  external symmetry breaking perturbations}
\label{subsec:nonsym_perturbation} In reality, a strictly
symmetric substrate, either spherical or cylindrical, is
impossible.
Also, the chemical properties of the substrate surface are never
perfectly homogeneous.
A nominally spherical latex bead may contain a weak local
deviation of the surface curvature and a weak heterogeneity of the
polymerization constant, $\bar{k}_{p}$, along the surface.
Thus we should suppose that there are a disturbances which break
{\it externally} the symmetry of the system, and modify
Eq.(\ref{eq:reac0}) or its linearized form
 Eq.(\ref{eq:linear}).
A legitimate question is, therefore, ``if and how the above
symmetry breaking {\it instability} plays a role?''
%
In short, the answer is that, despite these extrinsic factors, the
instability mechanism of symmetry breaking manifests itself in the
evolution of the gel's thickness, justifying our comparison with
experiments done in the above \S\S\ref{subsec:evolution}
We discuss it first in a formal manner, and then in the context of
the geometrical and chemical heterogeneities.
Within the linear approximation, the evolution equation
Eq.(\ref{eq:reac0}) is decomposed into the equation for each mode,
like Eqs.(\ref{eq:epsilon0}) or (\ref{eq:linear}). In the latter
equation the system's intrinsic heterogeneity may be represented
as a small but finite source term, $\epsilon_q^e$, \beq
\frac{d\epsilon_q(t)}{dt}= \frac{1}{\tau_q}( \epsilon_q(t) +
\epsilon_q^e). \label{eq:bp} \eeq
We can solve this equation with the initial condition $\epsilon_q
(0)=0$: \beq \epsilon_q(t) =\epsilon_q^e \tau
(e^{\frac{t}{\tau}}-1) \simeq
\left\{\begin{array}[c]{lc}\epsilon_q^e/\tau_q t, & \mbox{for } t<\tau \\
\epsilon_q^e  {e^{\frac{t}{\tau}}}, &    \mbox{for } t>\tau
\end{array} \right. .
\label{eq:eqt} \eeq This shows that, after a time $t\sim $ a few
$\tau$, the effect of the non-symmetric disturbance
is 
exponentially amplified ($e^{t/\tau}\gg 1$) by the instability
mechanism, while the direct effect of the source
is small in the sense that $\epsilon_q^e t\ll 1$ even for $t
\simeq$ few $\tau_q$.
In this way, the symmetry breaking mechanism manifests itself as
an amplifier of small heterogeneous disturbance in the system,
which can be experimentally observable.
%

An other way to think about the external perturbation is to define
the time $\tau_{Sq}$ required for developing an $\epsilon_q$  of a
specified value $\epsilon_q^S$: Equation (\ref{eq:eqt}) leads to:
\beq \tau_{Sq}=\tau_{q} \ln
\left(1+\frac{\epsilon_q^S}{\epsilon_q^e}\right). \label{eq:jp}
\eeq Changing the prescribed value $\epsilon_q^S$ or the external
perturbation $\epsilon_q^e$ by  orders of magnitude changes
$\tau_S$ only by a small factor. This tells us that, as already
announced the scaling of the characteristic observable times is
essentially given by $\tau_q$. It also tells us that the detailed
knowledge of the early dynamics is not essential in the definition
of $\tau_{Sq}$ provided $\tau_q$ is sufficiently larger than
$\tau_0$, so that a quasi spherical state is obtained before the
symmetry breaking process is observed. We know this to be true
both from our analysis and from experiment.

Now we describe how the parameter $\epsilon_q^e$ reflects the
effect of the heterogeneity of the surface curvature and of the
polymerization rate.
Under the linear approximation, we only need to consider the
profile of the substrate surface which can be described in terms
of the radius $r_0(\theta_0)$ as a function of the angle
$\theta_0$: $r_0(\theta_0)=r_{0}+\Delta_{q}^{\rm
(geo)}\cos(q\theta_0)$.
Additionally we consider the spatial distribution of the
polymerization rate constant $\bar{k}_p$ represented as a function
of $\theta_0$:
$\bar{k}_p(\theta_0)=\bar{k}_{p0}+\Delta_{q}^{(chem)}
\cos(q\theta_0)$. $\Delta_{q}^{\rm (geo)}$ and
$\Delta_{q}^{(chem)}$ characterize the amplitudes of geometrical
and chemical perturbations, respectively.
The geometric profile $r_0(\theta_0)$ leads to the non-homogeneous
curvature $\kappa(\theta_0)$, which has the following form,
\[ \kappa(\theta_0)
=\frac{1}{r_0}\left[1+\frac{\Delta_{q}^{\rm
(geo)}}{r_0}(q^2-1)\cos(q\theta_0) \right].\]
Along the line of calculation in Appendix D, this expression of
the curvature should replace the factor ${r_0}^{-1}$ in
Eq.(\ref{eq:radial-fin}). The normal stress on the substrate
surface is therefore given by
\beq \left.\sigma_{rr}\right|_{r_0} =-\kappa(\theta) T.
\label{eq:curve} \eeq
As for the chemical heterogeneity in the polymerization rate,
$\bar{k}_p(\theta_0)$ should replace $\bar{k}_{p}$ in
Eq.(\ref{eq:reac0}).
In general these effect can be summarized in the form of\\
$\epsilon_q^e =\beta_{q}^{(chem)}
({\Delta_{q}^{(chem)}}/{\bar{k}_p})+ \beta_{q}^{(geo)}
({\Delta_{q}^{(geo)}}/{r_0})$, with dimensionless numbers
$\beta_{q}^{(chem)}$ and $\beta_{q}^{(geo)}$.
However, if these source terms have existed from the start of
polymerization, the expressions of $\beta_{q}^{(chem)}$ and
$\beta_{q}^{(geo)}$ are complex because
in the early stages of the gel growth none of the linear equation
is valid.
However, as we have already pointed out the exact knowledge of
$\epsilon_q^e$ is not essential for understanding the main feature
of the dynamics if $\tau_0 < \tau_q$.
We, therefore, only mention about the restricted case where those
heterogeneities are switched on at a certain moment of time after
the symmetric steady state has been established. The result then
reads
\beq
\epsilon_q^e=
\frac{e^{c_d B\chi }}{\Omega_q \chi }\left(
\frac{\Delta_{q}^{(chem)}}{\bar{k}_p}-\frac{c_p B\chi
^2}{2}(q^2-1) \frac{\Delta_{q}^{\rm (geo)}}{r_0} \right).
\label{eq:bq} \eeq The positive coefficient in front of
$\Delta_{q}^{(chem)}$ reflects the acceleration of the turnover of
the gel material where $\bar{k}_p$ is increased, while
the minus sign in front of the second term in the bracket reflects
the polymerization being slowed down where the surface extrudes,
or, where $\kappa(\theta_0)>\frac{1}{r_0}$.


%

\section{Discussion}
\label{sec:discussion} Our analysis based on gel elasticity leads
to the essential prediction that the spherical symmetry is always
unstable.
The expected scenario is that in a first step a quasi-spherical
steady state is reached which should obey the prediction contained
in {\cite{noireaux2000} and \cite{julie2003}.
Then on a time scale significantly larger than the characteristic
time for reaching the isotropic quasi-steady state, symmetry is
broken.
In the regime we discuss, governed by elasticity, these two times
are predicted to scale like the radius of the bead on which the
experiment is conducted.
This scaling should be very robust in the elastic regime, since
$r_0$ is the only length scale in the problem.
In particular it should hold for wavelength larger than those
considered here.
All these expectations are well born out by experiment
\cite{anne2002}.
In a number of cases symmetry is not observed to be broken: this
may be due to three different causes:
- the experiment duration might not be long enough for the
symmetry breaking event to take place,
- the gel/bead friction, considered in appendix E might further
slow down the symmetry breaking process,
- the gel might not behave fully elastically at very long time
scales.
In this latter case, a new time scale would come into play, namely
that over which a significant stress may be maintained, and a new
calculation should be developed.
We discuss various possible improvements to our current analysis
in appendix E.

As explained in this manuscript the main ingredient for the
occurrence of symmetry breaking comes from the tensile stress
concentration where the gel thickness is smallest.
This feature, added to a stress dependent depolymerization in the
immediate vicinity of the gel outer surface, leads to an absolute
instability of the system.
This is in contrast with earlier models
{\cite{oudenaarden1999,mogilner2003}} in which symmetry is broken
at polymerizing gel bead surface.
Their interpretation is most transparent in the one dimensional
case; consider two opposing sides on which parallel filaments are
grown.
The force on individual filaments, i.e. the ratio of the total
force (equal on both sides because of force balance) to the number
of supporting filaments, is the key notion.
The smaller the number of filaments participating, the slower the
effective polymerization rate;
it is natural to expect a force dependence and different scenarios
have been discussed \cite{oudenaarden1999,mogilner2003} if an
unbalance between the two sides arises at some point, it grows
since the ``weak'' side tends to become ``weaker''.
The two-dimensional version of this mechanism, simulated by van
Oudenaarden {\it et al.} \cite{oudenaarden1999} is closely related
to simulation and experiments done on the microtubule/centrosome
(or microtubule/bead) system. The latter system does not exhibit
an instability whereas the first does. The difference in behavior
results from difference in boundary conditions. All these cases do
not consider the situation where filaments are crosslinked.
%
%
%
Actin gels are crosslinked and we propose that in two and three
dimensions these crosslinks change profoundly the behavior.
%
%
%
%
Indeed,
 if the thickness of the gel layer is locally
decreased, the compressive stress there,
$\left.\sigma_{rr}\right|_{r_0}$, should either stay constant if
full-slip boundary conditions are achieved, or decrease,
irrespective of the thinning cause.
The lateral displacement of the gel layer along the substrate
surface might at most relax some of this local decrease of
$\left.\sigma_{rr}\right|_{r_0}$, but it will never be able to
increase it.
%
%
%
%
%
Thus the mechanism described \cite{oudenaarden1999,mogilner2003}
for non-crosslinked filaments do not apply to gels.

A direct experimental assessment of the symmetry breaking
mechanism could involve monitoring simultaneously the
depolymerization and the polymerization processes at the outer and
inner gel surfaces, respectively.
This is not an easy experiment.
%

%

\section*{Acknowledgement}
We thank M.-F. Carlier for the gift of the medium of motility. We
also thank C.Sykes for fruitful discussions and for critical
reading of the manuscript.


\clearpage
\section*{Appendix A. Heuristic derivation of the equations of
mechanical balance}

Eqs. (\ref{eq:bl3}) and (\ref{eq:bl2}) in \S~\ref{sec:symmetric}
are the equation of mechanical balance of stress components in the
spherically and circularly symmetric geometries, expressed in
respective relevant coordinate systems.
Instead of deriving these from the familiar form in the Cartesian
coordinates (symbolically written as $\nabla \cdot \sigma =0$)
through coordinate transformations, we will present an elementary
physical interpretation of the equations of mechanical balance.
It might help to understand how the lateral tensile stress and the
normal compressive stress are related. See, {\bf Fig.\ref{fig:6}}.
\begin{figure} 
\vspace{5cm} %
\caption{Forces acting on a curved slice of gel of thickness
$\Delta r$ and length $r\Delta \theta$. The slice is under lateral
tension because of forces $\sigma_{\perp\perp} \Delta r$. It is
radially compressed because of the forces $(r+\Delta r)\Delta
\theta \left.\sigma_{rr}\right|_{r+\Delta r}$ and
$r\Delta \theta \left.\sigma_{rr}\right|_{r}$.} %
\label{fig:6} %
\end{figure} %
Consider, within a layer of actin gel occupying the radii $r_0$
and $r_0+h$, a slice of gel between the radii $r$ and $r+\Delta r$
spanning a solid angle $\Delta \Omega$ (3D) or an angle $\theta$
(2D).
The lateral tension $\sigma_{\perp\perp}$ gives an effective
 surface tension
$\Delta \Gamma=\sigma_{\perp\perp} \Delta r$ to this slice.
Because of the curvature radius, $r$, of this slice, a sort of the
Laplace pressure,  $2\Delta \Gamma/r$ (3D) or $\Delta \Gamma/r$
(2D) is generated towards the center ($r=0$) of curvature.
This pressure integrated over the surface, $r^2 \Delta \Omega$
(3D) or $r \Delta \theta$ (2D), gives the total force exerted by
this thin layer.
These force in the respective dimensionality are shown on the
right hand side of the equations below.
Now, these forces must be the origin of the difference between the
integrated normal stress acting at $r$ and that at $r+\Delta r$.
The differences in the respective dimensionality are shown on the
left hand side of the equations below.
\[  (r+\Delta r)^2 \Delta \Omega \,\sigma_{rr}|_{r+\Delta r}
 -  r^2 \Delta \Omega \, \sigma_{rr}|_{r}
=r^2  \Delta \Omega \, \frac{2 \sigma_{\perp\perp} \Delta r}{r} ,
\! \mbox{(3D)}\]
\[
(r+\Delta r) \Delta \theta \, \sigma_{rr}|_{r+\Delta r}
 -  r \Delta \theta \, \sigma_{rr}|_{r}
=  r  \Delta \theta \,\frac{ \sigma_{\perp\perp} \Delta r}{r} , \!
\mbox{(2D)}\]
Dividing the both hand sides of the above equations by $\Delta r$,
and letting $\Delta r\to 0$, we arrive at Eqs. (\ref{eq:bl3}) and
(\ref{eq:bl2}).
Note that $\sigma_{rr}<0$ for compressive stresses.

\section*{Appendix B: Elementary physical mechanism
of the stress concentration}
We describe the basic mechanism of the stress concentration by an
illustrating example with a very simple geometry (see, {\bf
Fig.\ref{fig:7}(a)}).
\begin{figure} 
\vspace{5cm} %
\caption{Stress distribution in elastic structures of varying
thickness: (a) Elastic rod of thickness $d(z)$ under tension due
to forces $F$ acting at the ends. The tension at coordinate $z$:
$\sigma_{zz}\simeq F/A(d)$, where $A(d)=\frac{\pi}{4} d(z)^2$. (b)
Analogous situation in a gel layer of thickness $h(\theta_0)$
under integrated tension $T$. The tensile stress can be
approximated as
$\sigma_{\perp\perp} \simeq \frac{T}{h(\theta)}$.} %
\label{fig:7} %
\end{figure} %

Suppose that there is a long elastic rod whose diameter $d(z)$ is
inhomogeneous along its long axis, $z$.
We now apply a tensile force to this rod by pulling its ends
apart.
Once the balance of force is reestablished within the rod, the
total tensile force integrated over a sectional plane
perpendicular to the $z$-axis is constant along of the
$z$-coordinate.
Thus the tensile stress $\sigma_{zz}$ averaged over this section
is inversely proportional to its area, $\pi (d(z)/2)^2$.
By such geometrical effect, the tensile stress is concentrated at
the thinnest part of the rod.

We could mention a analogous situation in an electric wire
transporting a steady electric current.
If the thickness of the wire is inhomogeneous, the electronic
current {\it density} is high in the region where the wire is
thin, by the same geometrical effect.
The electric current density plays the role of the tensile stress
$\sigma_{zz}$ in the former case.
In fact, the stress is the current density of the momentum
\cite{landau1967}.

We may compare these quasi one dimensional examples with the
geometry studied in \S\ref{sec:concentration} (see {\bf
Fig.\ref{fig:7}}(b)).
In the latter situation, the gel is under lateral tensile force
$T$.
By the same reasoning as above, the lateral tensile stress is
large where the thickness is small.
Although the shear force between the gel and the cylinder would
weaken this effect, the basic mechanism still works.

The geometrical effect discussed here is quite universal:
We only need a current density of some physical quantity (ex. the
momentum, the charge, etc.) which is confined along some
direction(s).
%

\section*{Appendix C}
\label{sec:appendixC}

We demonstrate that the perturbations of the surface profile have
a limited influence on the stress, practically confined within a
``skin depth'' near the outer surface of the gel layer, where the
skin depth is of the order of the wavelength of
the perturbation (see Fig.\ref{fig:3} in the text). \\
{\it (i)} To the spherically symmetric gel layer, we introduce a
$xyz$-coordinate system so that its $xy$-coordinate plane is
tangent to the outer surface of the gel layer at its origin,
$x=y=z=0$.
We define the sign of $z$-coordinate so that the bulk of the gel
is on the side of $z\le 0$.
We will consider a small neighborhood of the origin so that the
curvature of the gel surface is negligible.
This apparently flat gel layer is under lateral
tension along the $xy$-plane.\\
{\it (ii)} We introduce a slight sinusoidal perturbation of the
surface profile of the gel layer, without allowing the
displacement of the gel material.
The perturbed surface profile is written as\\
$z=$ $\epsilon_Q$
$Re[e^{iQx+\phi}]$, with the amplitude $\epsilon_Q$, the wave
number $Q$ and the phase
$\phi$ being constant.\\
{\it (iii)} We then let the gel to relax until the mechanical
balance is reestablished within the layer.
By this process the stress components $\sigma_{\alpha\beta}$ with
$\alpha=x$, $y$, or $z$ are also perturbed.
We denote by $\delta \sigma_{\alpha\beta}$ the perturbed part of
the stress components.
These $\delta \sigma_{\alpha\beta}$ obeys the following equations:
\[
\frac{\partial}{\partial x}\delta \sigma_{x\alpha}+
\frac{\partial}{\partial y}\delta \sigma_{y\alpha}+
\frac{\partial}{\partial z}\delta \sigma_{z\alpha}=0,
\]
with $\alpha=x$, $y$, or $z$
We assume that the usual linear elasticity relationship applies to
the system. Then the perturbed stress components is related with
the displacements $(u_x, u_y, u_z)$ from the unperturbed state
through the equation:
\[ \delta \sigma_{\alpha\beta}=2\mu(u_{\alpha\beta}
-\frac{1}{3}\sum_{\gamma}u_{\gamma\gamma} \delta_{\alpha\beta})
+K\sum_{\gamma}u_{\gamma\gamma}{\delta_{\alpha\beta}},\] with
\[u_{\alpha\beta}\equiv
\frac{1}{2}(\frac{\partial u_\alpha}{\partial x_\beta}+
\frac{\partial u_\beta}{\partial x_\alpha}),\] where $\mu$ and $K$
are the shear and bulk moduli, the suffices $\alpha$, $\beta$ and
$\gamma$ take $x$, $y$ or $z$, and $\{x_x,x_y,x_z\}\equiv
\{x,y,z\}$. The summation index $\gamma$ runs over $x$, $y$ and
$z$, and
$\delta_{\alpha\beta}$ is the Kronecker's delta.\\
{\it (iv)}
The question is how the quantities $\delta \sigma_{\alpha\beta}$
depend on $z$ for $z<0$.
In the lowest order of $\epsilon_Q$, $\delta
\sigma_{\alpha\beta}$, and therefore the displacements $u_\alpha$
should depend on $x$ sinusoidally with the wavenumber, $Q$.
Therefore, the above equations can be reduced to the following
matrix equation: {\small
\[
\left(
\begin{array}[c]{c}
 2\mu \frac{\partial^2}{\partial z^2}-(K+\frac{10}{3}\mu)Q^2
                  \quad  \qquad       \qquad iQ \frac{\partial}{\partial z}   \\
 (K+\frac{4}{3}\mu)iQ\frac{\partial}{\partial z}
                           \qquad  (K+\frac{10}{3}\mu)\frac{\partial^2}{\partial z^2}- 2\mu Q^2
\end{array}
\right) \left(\! \begin{array}[c]{cc}  u_x \\ u_z  \end{array}\!
\right) \!=\! \left(\! \begin{array}[c]{cc}  0 \\ 0
\end{array}\!\right).
\]
}
This equation can be finally reduced to the following equations:
\[ ({\partial_z}^2-Q^2)^2 \psi  =0 \]
where $\psi$ is a combination of $u_x$ and $u_z$.
From this equation, we find that the displacements should depend
exponentially on $z$.
Among mathematically possible forms $e^{\pm Qz}$, we discard the
form $e^{-Qz}$ since this factor grows exponentially towards
negative $z$ axis.
We are then left with the form $e^{Qz}$ for $z<0$.
This indicates that the influence of the perturbations to the
surface profile with the wavelength $\sim Q^{-1}$ is practically
limited within a region with a``skin depth'' $\sim Q^{-1}$ from
the outer surface of the gel layer.

\section*{Appendix D. The derivation of
Eqs.(\protect\ref{eq:lateral-lin}) and (\protect\ref{eq:sigma_rr})
 } \label{sec:appendixD}

Here we show how the stress distribution within the gel layer is
calculated for the model described by {\it (a)}-{\it (c)} in
\S\ref{sec:concentration}.
In the polar coordinate,
the equations of mechanical balance 
in the gel layer is written as follows:
\begin{eqnarray}
    \frac{\partial}{\partial r}(r\sigma_{rr})
+\frac{\partial}{\partial \theta} \sigma_{r\theta}
    -\sigma_{\theta\theta} & = & 0,\nonumber\\
     \frac{\partial}{\partial r}(r\sigma_{r\theta})
+\frac{\partial}{\partial \theta} \sigma_{\theta\theta}
    +\sigma_{r\theta} & = & 0.
\end{eqnarray}
We integrate the left hand side of these equations with respect to
$r$ from $r_0$ to $r_0+h$, noting the boundary conditions
Eqs.(\ref{eq:cisaisub}) and (\ref{eq:stressfree}).
%
The result reads
\[
 T-\frac{\partial}{\partial \theta}\bar{T}
= -r_{0}\sigma_{rr}\vert_{r_{0}}
\]
and
\[
 \bar{T}+\frac{\partial}{\partial \theta}T
=0,
\]
where we have introduced the total tension,
$T\equiv \int_{{r_0}}^{{r_0}+h} \sigma_{\theta\theta}dr,$ and its
analogue for the shear stress, $\bar{T}\equiv
\int^{r_0+h}_{r_0}\sigma_{r\theta}\,dr$.
Based on our estimates of the shear stress, $\sigma_{r\theta}\sim
\tilde{\mu}\epsilon_q {(h^*\!/r_0)}^2 $, and the perturbed part of
the tensile stress, $\delta\sigma_{\theta\theta}\sim \epsilon_q B
h^*\!/r_0$ (see \S\ref{sec:concentration}), we can evaluate the
terms on the left hand side of the above equations. Since
$\frac{\partial}{\partial \theta}\sim 1$ under our limitation of
the wavenumber $q$, we have, $\delta T\sim
\frac{\partial}{\partial \theta}T \sim B \epsilon_q {h^*}^2/r_0$
and $ \bar{T} \sim \frac{\partial}{\partial \theta}\bar{T} \sim
\tilde{\mu} \epsilon_q {h^*}^3/{r_0}^2,$ where $\delta T$ is the
perturbed part of the total tension $T$. Assuming $\tilde{\mu}\sim
B$, we find that $\bar{T}$ and $\frac{\partial}{\partial
\theta}\bar{T}$ are smaller than  $T$ and
$\frac{\partial}{\partial \theta}T$ by a factor of $h^*/r_0$.
We, therefore, ignore the terms with $\bar{T}$, and have the
following equations:
%
\beq
   \frac{\partial}{\partial \theta}T =0, \quad
%
    T =-r_{0}\sigma_{rr}\vert_{r_{0}}.
\label{eq:T-pascal} \eeq
The first equation 
requires the lateral balance of the integrated tension $T$,
while the second equation 
requires the homogeneity of the normal compressive stress on the
substrate surface, $\sigma_{rr}\vert_{r_{0}}$.

From (\ref{eq:T-pascal}) we can calculate
$\left.\sigma_{\theta\theta}\right|_{r_0+h^*}$ and $\left.
-\sigma_{rr} \right|_{{r_0}}$.
We employ the ``stacked rubber band model''
\cite{noireaux2000,gerbal2000,landau1967} for the lateral tensile
stress $\sigma_{\theta\theta}$, as we did for the symmetric case
(see Eq.(\ref{eq:rubber}) in the text).
Here we take into account the possible lateral displacement of the
gel layer upon the reestablishment of the mechanical balance.
We introduce an unknown function $\theta(\theta_0)$ such that the
material of gel layer originally at $\theta_0$ is moved to
$\theta(\theta_0)$ upon the reestablishment of the mechanical
balance (see {\bf Fig.\ref{fig:8}}).
\begin{figure} 
\vspace{5cm} %
\caption{(a) $\theta_0$ is defined as the angle with respect to a
reference line, when a material point (for example, the black dot)
is located before the gel is deformed. (b) as a result of elastic
deformation, the material point characterized by $\theta_0$ is
displaced to a new position at $\theta$. The function
$\theta(\theta_0)$ characterizes
the elastic deformation.} %
\label{fig:8} %
\end{figure} %
The elongation ratio, $(r-r_0)/r_0$ in the Eq.(\ref{eq:rubber})
is, therefore, replaced by the form which dependents on the
parameter $\theta_0$: $({rd \theta(\theta_0)-r_0 d \theta_0})/{r_0
d \theta_0}$. Thus the lateral tension is written as
\beq
 \sigma_{\theta\theta}  = B\,\left(
\frac{r}{r_0} \frac{d \theta(\theta_0)}{d \theta_0}-1 \right),
\label{eq:lateral-bare} \eeq
where, the shear deformation within the layer has been
consistently ignored. (The justification of this approximation
concerning another source of error will be discussed in the
Discussion section,
see Appendix E.)%
With the definition of $T$ given above, we obtain
\beq T= {B} \left[\left(h(\theta_0) +\frac{h(\theta_0)^2}{2
r_0}\right) \frac{d \theta(\theta_0)}{d \theta_0} - h(\theta_0)
\right]. \label{eq:T-bare} \eeq
The function $\theta(\theta_0)$ can be related to $h(\theta_0)$
through the first equation in (\ref{eq:T-pascal}), which requires
that $T$ is constant.
To fix the value of the constant, $T$, we recall an apparent
condition $\int_0^{2\pi} \frac{d \theta(\theta_0)}{d \theta_0}
d\theta_0 = 2\pi$.
The result of $T$ is
\beqa
{T}&=& {B} {\left[{\int^{2 \pi}_{0}
\frac{d\theta_0}{h(\theta_0)+\frac{h(\theta_0)^2}{2 r_0}}
}\right]}^{-1}\,\cr
 && \left[{ 2\pi-\int^{2 \pi}_{0}
\frac{h(\theta_0)}{h(\theta_0)+\frac{h(\theta_0)^2}{2
r_0}}d\theta_0 }\right]. \label{eq:T-fin}
\eeqa
$\frac{d \theta(\theta_0)}{d \theta_0}$ can thus be finally
determined in terms of $h(\theta_0)$ (which we do not show
explicitly).
From the second equation of (\ref{eq:T-pascal}) and from
Eq.(\ref{eq:lateral-bare}), we have
\beqa  %
 \left.\sigma_{\theta\theta}\right|_{r_0+h(\theta_0)}  &=&
{\left[{h(\theta_0)+\frac{h(\theta_0)^2}{2r_0}}\right]}^{-1}\, \cr
&&
\hspace{-1cm}
\left[{{T}\,\left(1+\frac{h(\theta_0)}{r_0}\right)+
{B}\frac{h(\theta_0)^2}{2r_0}}\right], \label{eq:lateral-fin}
\eeqa %
\beq -\left. \sigma_{rr}\right|_{{r_0}}=\frac{T}{r_0}.
\label{eq:radial-fin} \eeq
To reach the expressions Eqs.(\ref{eq:lateral-lin}) and
(\ref{eq:sigma_rr}) in the text, we may simply substitute the form
$ h(\theta_0)= h^*\left[1 + \epsilon_q \cos (q \,\theta_0
)\right], $
into Eqs.(\ref{eq:T-fin})-
(\ref{eq:lateral-fin}), and develop them with respect to
$\epsilon_q$ up to the linear order.
%
%


\section*{Appendix E} 

Below we  mention briefly  aspects which could be improved in our
present analysis:


{\it 1. Extend the analysis to modes with $q \neq \frac{2\pi
r_0}{h^*}$:
}\\
Although the instability against the disturbances of modes $q
\simeq \frac{2\pi r_0}{h^*}$ is sufficient to destroy the system's
stability, our analysis can say nothing about what is the fastest,
or the most unstable, mode of the disturbance.
The fact that the characteristic time of the symmetry breaking in
our analyses gives reasonable values suggests that the other modes
of perturbations might grow, if they do, at a rate not highly
exceeding the one we have analyzed.
In fact, some efforts to refine the present analysis (see below)
indicate that, for $q < \frac{2\pi r_0}{h^*}$, the instability is
weakened or even suppressed, while for the modes, $q > \frac{2\pi
r_0}{h^*}$, there is no sign of appreciable $q$ dependences.
However, the modulation of the micro-scale comparable to the mesh
size of the gel is not accessible by the continuum approach.
(Note that the quasi one-dimensional analysis of
\cite{mogilner2003} claims the instability of the mode $q=2$.)

{\it 2. Remove the full slip boundary condition:}\\
As discussed in \S\ref{sec:concentration} we have justified this
boundary condition when we analyzed the evolution of the modes $q
\simeq \frac{2\pi r_0}{h^*}$, since, there, the choice of boundary
condition on the substrate surface is expected to be insensitive
to the stability result.
For the other modes, especially for $q < \frac{2\pi r_0}{h^*}$, we
should take into account the friction on this surface due to the
temporal linkage between the actin filament with the substrate
(\cite{gerbal2000}).
(About the discussion of the relation between the friction and the
temporal linkage, see \cite{tawada1991,gerbal1999}.)
As a modification of the present model, we have incorporated the
finite friction force on the substrate surface, which is
proportional to the slipping velocity of the gel along the
surface. Though details will not be shown \cite{sekimoto2001}. the
result indicate that, while the all modes remain unstable, the
instability is weakened for long wavelengths, i.e. for small
values of $q$.

{\it 3. Extend the analysis to non-linear regime:}\\
Our analysis does not infer how the comet of actin gel is formed
and continues to grow after the symmetric shape of the layer
around the bead is lost.
This is a nonlinear problem.
\cite{oudenaarden1999} have demonstrated in their numerical
modelling that the comet formation shows its optimal performances
for a certain parameter value related to the depolymerization at
the substrate surface.
The comparative study from our point of view is yet to be done.
(As for the steady growth of the comet from Listeria, see
(\cite{gerbal2000})).

{\it 4. Extend the analysis to soft beads:}\\
Endosomes, lysosomes, vesicles and fluid drops deform as the comet
develops, revealing the importance of mechanical stresses
\cite{taunton2000,theriot2003,oudenaarden2003}.
The deformation of a fluid drop has been fully analyzed within the
framework of the elastic analysis and shown to be quantitatively
in agreement with the experiment\cite{campas200x}.
The symmetry breaking onset remains to be worked out.
As proposed in the discussion, the early stages would discriminate
between the different mechanisms.

{\it 5. Removal of the assumption of the isotropic gel:}\\
Generally speaking, the micro-structure of the gel polymerized
from a surface must distinguish the radial direction from the
lateral ones.
Especially, for the actin gel branched by the help of the protein
Arp2/3 is shown to have a topology like a ``forest'' rather than
the network \cite{cameron2001}.
It will be the entanglement among the branches of the ``trees'' of
semi-flexible filaments that supports the tensile stress within
the gel.
Though we expect no qualitative change of our result upon the
incorporation of the elastic anisotropy of the gel, there should
be quantitative differences.
For the further analysis, we also need the experimental data on
the anisotropic elastic constants \cite{landau1967}.

{\it 6. Extend the analysis where the gel density is spatially
heterogeneous:
}\\
The effect of the spatial heterogeneity of the catalytic activity
of enzyme may have several aspects.
The one which has been discussed in
\S\S\ref{subsec:nonsym_perturbation} is the the modulation of the
polymerization rate, $\bar{k}_p$.
The other aspect which is related to the spatial heterogeneity of
elastic moduli of gel may also deserve consideration.
In fact, the heterogeneity of the elastic moduli will be closely
related to the heterogeneity of the factors $k_d$ and $c_d$ both
concerning the depolymerization processes kinematically and
energetically, respectively.
It is therefore impossible to predict where does the thinning of
the gel layer proceed most rapidly.
However, the rule of thumbs is again that the positive feedback
mechanism mentioned above: once the degradation is advanced in a
portion of gel layer than elsewhere, the stress concentration is
most likely to occur and the degradation will further accelerated
there.
Visco-elastic or frictional effects in the bulk gel or on the
substrate surface, respectively, may limit this positive feedback
loop. The detailed discussion will be the task of future works.

{\it 7. Take account of the frustration of stress in gel:
}\\
In the analysis of \S\ref{sec:mechano_chem}, the total tension
$T\equiv \int_{{r_0}}^{{r_0}+h} \sigma_{\theta\theta}dr$ has been
calculated by substituting the expression of the stacked rubber
band model, Eq.(\ref{eq:lateral-bare}).
This operation ignores the fact that the gel material at two
different radii are created at different points of time.
The Fig.\ref{fig:5} illustrates how
 the simultaneous polymerization
and lateral deformation create a mechanical frustration within the
gel material:
\begin{figure} 
\vspace{5cm} %
\caption{Generation of a mechanical frustration caused by
simultaneous polymerization and lateral deformation a rectanglar
piece of gel (marked block) is displaced and deformed while
keeping the connectivity with its neighboring piece (block shown
by dotted lines).} %
\label{fig:5} %
\end{figure} %
(a) Consider a thin slice of actin gel created at the substrate
surface  during a short time interval, say, between $t_1$ and
$t_1+dt$ (the dark gray region occupying the angle $\Delta
\theta_0$).
We may expect that this part of gel which is just grown bears no
lateral stress, to a good approximation.
In the context concerning this slice just above the substrate
surface, we would then set $d\theta(\theta_0)/d\theta_0=1$ in
Eq.(\ref{eq:lateral-bare}).
(b) After the consecutive time interval, $t_1+dt<t<t_1+2dt$, the
same spatial region, which is now indicated by the dotted lines,
is occupied by a newly grown gel under no lateral tension. Thus
again $d\theta(\theta_0)/d\theta_0=1$ for this region.
However, as for the previously grown material which we have marked
in dark gray in (a), it now occupies the region just outside the
original one (shown again in dark gray), and occupies the angle
$\Delta \theta$.
Generally,  $\Delta \theta$ is different from $\Delta \theta_0$ as
far as there is a global lateral displacement of gel during the
time interval $t_1+dt<t<t_1+2dt$.
Thus in the context concerning this region in dark gray, we would
set $d\theta(\theta_0)/d\theta_0\neq 1$ in
Eq.(\ref{eq:lateral-bare}).
This contradiction indicates a natural process through which a
mechanical frustration is created within the gel layer, and shows
that Eq.(\ref{eq:lateral-bare}) is only approximative.

%
Taking account of this fact in the model
requires a lot of complication of the formalism, but the linear
analysis is still feasible. Though the details will not be shown
\cite{sekimoto2001}, the result indicates that, while the modes
with small $q$ values now become stable, the instability persists
for $q>q_c$ with a finite positive threshold $q_c$.
Our simple analysis with Eq.(\ref{eq:lateral-bare}) is still a
good approximation if the characteristic time of the instability
$\tau$ is short enough as compared with the turnover time of the
gel, $r_0/v_{\rm gel}$.
From Eq.(\ref{eq:taur0}) this criterion
 reads $(2+\chi)/(c_d B\chi)\ll 1$.
Substituting the same values for $c_d=\xi^3/T$, $B$, and $\chi$ as
in \S\S\ref{subsec:evolution}, the left hand side of the above
criterion becomes
 0.16 if we take $\xi=30$nm for the mesh size of
the actin gel.
We, therefore, suppose that our approximation is pretty good for
the above parameter range.
%

%

%

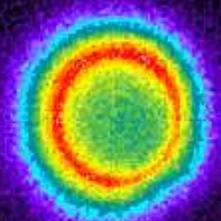
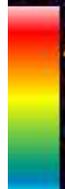
\end{document}